\documentclass[showpacs,twocolumn,prl]{revtex4-1}
\usepackage[colorlinks, hyperfigures, linkcolor=black, citecolor=black, breaklinks,]{hyperref}
\usepackage[centertags]{amsmath}
\usepackage{amssymb,eqnarray}
\usepackage{amsthm}
\usepackage{fancyhdr}
\usepackage{graphicx,subfigure}
\usepackage{epsfig}
\usepackage{epsfig}
\usepackage{amssymb}
\usepackage{bbm}
\usepackage{graphicx}
\usepackage{amsmath}
\usepackage{subfigure}

\newcommand{\p}{\boldsymbol{p}}

\newcommand{\rr}{\boldsymbol{r}}

\newcommand{\xxi}{\boldsymbol{\xi}}
\newcommand{\xchi}{\boldsymbol{\chi}}
\newcommand{\eeta}{\boldsymbol{\eta}}

\newcommand{\sn}[1]{{\color{black}}}

\begin{document}
	\title{Pseudochemotaxis in inhomogeneous active Brownian systems}
	\author{Hidde D. Vuijk\textsuperscript{1}, Abhinav Sharma\textsuperscript{1}, Debasish Mondal\textsuperscript{1,2}, Jens-Uwe Sommer\textsuperscript{1,3}, Holger Merlitz\textsuperscript{1}}
	\address{\textsuperscript{1}Leibniz-Institut f\"{u}r Polymerforschung Dresden,
		Institut Theorie der Polymere, 01069 Dresden, Germany\\
		\textsuperscript{2} Indian Institute of Technology, Department of Chemistry, 517506 Tirupati, India\\
		\textsuperscript{3} Technische Universit\"{a}t Dresden, Institute of Theoretical Physics, 01069 Dresden, Germany 
		}

\begin{abstract}

We study dynamical properties of confined, self-propelled Brownian particles in an inhomogeneous activity profile. Using Brownian dynamics simulations, we calculate the probability to reach a fixed target and the mean first passage time to the target of an active particle. We show that both these quantities are strongly influenced by the inhomogeneous activity. When the activity is distributed such that high-activity zone is located between the target and the starting location, the target finding probability is increased and the passage time is decreased in comparison to a uniformly active system. Moreover, for a continuously distributed profile, the activity gradient results in a drift of active particle up the gradient bearing resemblance to chemotaxis. Integrating out the orientational degrees of freedom, we derive an approximate Fokker-Planck equation and show that the theoretical predictions are in very good agreement with the Brownian dynamics simulations.

		
\end{abstract}
	\maketitle
\section{Introduction}
	Active matter is ubiquitous in Biology. Examples include cytoskeletal molecular motors performing directed motion on filaments inside a cell \cite{julicher1997modeling}, nucleic acid motors involved in transcription process inside nucleus \cite{singleton2007structure}, and even microscopic living objects such as the \emph{Escherichia coli} bacteria which generates motion using helical flagella \cite{berg2008coli}. The defining characteristic of active matter is that it is intrinsically nonequilibrium. The constituents of active matter generate motion by consuming energy from their local environment. In addition to the solvent-induced Brownian motion, active particles undergo self-propulsion resulting in persistent character of particle trajectories. It is the self-propulsion feature of these particles, generally termed as the activity, which captures the nonequilibrium nature of active matter.

	
There exist synthetic~\cite{saha2014clusters,hong2007chemotaxis,magiera2015trapping,stenhammar2016light} and living systems~\cite{berg1972chemotaxis,jekely2008mechanism,hoff2009prokaryotic,jikeli2015sperm} 
for which the activity is not uniform but dependent on the spatial location 
of the particles. Theoretical studies of inhomogeneous active systems predict some very interesting phenomena such as directional transport of colloids inside a bath of active walkers \cite{merlitz2017directional}, torque-free polarization of active Brownian particles \cite{sharma2017brownian}, pseudochemotaxis \cite{ghosh2015pseudochemotactic} in an activity profile and activity density waves \cite{geiseler2016chemotaxis}.
 Inhomogeneous activity is a fundamental feature of self-phoretic active particles which exhibit highly interesting pattern formation and dynamics~\cite{liebchen2017phoretic,pohl2014dynamic,liebchen2015clustering,meyer2014active,sengupta2009dynamics} . These particles exhibit autochemotactic behavior in the sense that their motion is governed by the gradients of chemicals they produce. Position-dependent activity also features in the energy depot model~\cite{schweitzer1998complex}. However, it is only recently that position-dependent activity is realized in experiments. Using an inhomogenous laser field~\cite{lozano2016phototaxis}, synthetic microswimmers drifted towards increasing laser intensity exhibiting phototaxis. In a very recent experimental development, even an archaic form of chemotaxis has been observed: drift of self-propelled supramolecular motors towards increasing fuel concentrations~\cite{peng2015self}.


The previous studies pose a fundamental question: Is active motion directly related with chemotaxis by physical principles?  A first step towards an answer is the inspection of a most simple system which features both activity and a very primitive concept for a "food-source" namely an increasing activity towards the source caused by a gradient of available food in the proximity of the source. Motivated by these considerations, we consider in this work systems for which the activity varies in space, focusing on two dynamical properties of an active system: target finding probability and the mean first passage time (MFPT) to target. The target finding probability is the probability that a particle introduced at a given location, exits through a specified boundary representing the target. The mean first passage time to target is a measure of average reaction time in finite domains \cite{szabo1980first}.

\begin{figure}[t]
\includegraphics[width=0.7\columnwidth]{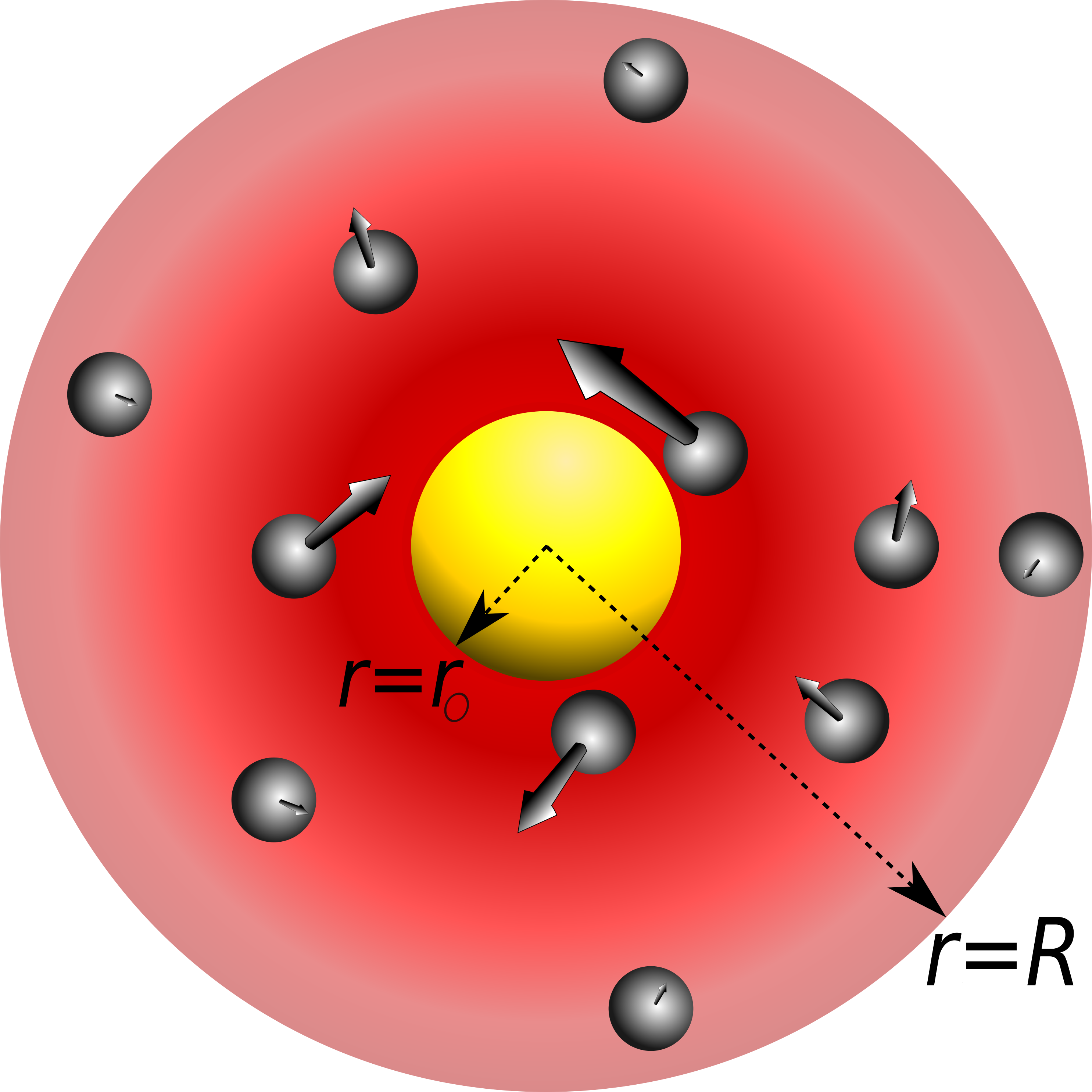}
\caption{Schematic of the inhomogeneous active system. The surface of the yellow spherical shell $r=r_0$ represents the target, for instance fuel or energy source for active particles. The fuel distribution in the spherical geometry is shown in red, the gradient of which represents the decaying activity profile away from the source. Active particles are shown as gray-white spheres together with their embedded orientation vector. The size of the vector represents the local self-propulsion speed of the particle. 
%
} 
\label{schematic}
\end{figure}

  In the context of active systems, the average reaction time has been previously studied in homogeneous systems \cite{sharma2017escape, scacchi2017mean}. Here we show that both these dynamical properties are strongly dependent on the spatial distribution of activity. In particular, we find that the insights gained from steady-state inhomogeneous active systems, such as preferential accumulation in the low-activity regions and orientation of particles antiparallel to the activity gradient, cannot be used to understand the dynamical properties of an inhomogeneous active system. We show that these quantities are strongly influenced by the inhomogeneous activity. When the activity is distributed such that high-activity zone is located between the target and the starting location, the target finding probability is increased and the passage time is decreased in comparison to a uniformly active system. Moreover, for a continuously distributed profile, the activity gradient results in a drift of active particle up the gradient bearing resemblance to chemotaxis. Our theoretical predictions are based on an approximate Fokker-Planck equation and are shown to be in very good agreement with the Brownian dynamics simulations.

  
  %


\section{Model and Theory}\label{modelandtheory}

We consider a three dimensional system of active, noninteracting, spherical Brownian particles 
with position $\rr$ and orientation specified by an embedded unit vector $\p$ (see Fig. \ref{schematic}). 
A space-dependent self-propulsion speed $v_0(\rr)$ 
acts in the direction of orientation. 
Omitting hydrodynamic interactions the motion can be modeled by the Langevin equations
\begin{align}\label{full_langevin}
&\!\!\!\!\!\!\dot{\rr} = v_0(\rr)\,\p  + \xxi\;\;,
\;\;\;
\dot{\p} = \eeta\times\p \,.
\end{align}
%
The stochastic vectors $\xxi(t)$ and $\eeta(t)$ are Gaussian distributed with zero mean and 
have time correlations	
$\langle\xxi(t)\xxi(t')\rangle=2D_t\boldsymbol{1}\delta(t-t')$ and 
$\langle\eeta(t)\eeta(t')\rangle=2D_r\boldsymbol{1}\delta(t-t')$. 
The translational and rotational diffusion coefficients, $D_t$ and $D_r$, are treated 
as independent parameters. 
The set of equations in \eqref{full_langevin} are convenient for Brownian dynamics simulations. However, on averaging out the orientational degrees of freedom, one obtains a theoretically tractable model of active particles evolving according to the Langevin equations~\cite{fily2012athermal,farage2015effective,wittmann2016active,maggi2015multidimensional,marconi2015}
\begin{align}\label{integrated_langevin}
&\!\!\!\!\!\!\dot{\rr} = \xxi + \xchi(\rr)\,.
\end{align}
Here, the stochastic force $\xchi(\rr)$ is position-dependent and has the time correlation
$\langle\xchi(\rr,t)\xchi(\rr,t')\rangle=D_\text{a}(\rr)\boldsymbol{1}\tau_\text{a}^{-1} e^{-|t-t'|/\tau_\text{a}}
 $,
 where $D_\text{a}(\rr)=v_0^2(\rr)\tau_\text{a}/3$ denotes a position-dependent coefficient and $\tau_\text{a}=(2D_\text{r})^{-1}$ is the persistence time of the orientation of the active particle. Due to the presence of colored noise in Eq.~\eqref{integrated_langevin}, an exact Fokker-Planck equation for the time evolution of probability density cannot be obtained. 
 Here, we use the simplest approximation of reducing $\xchi(\rr)$ to a white noise with time correlation $\langle\xchi(\rr,t)\xchi(\rr,t')\rangle=2D_\text{a}(\rr) \boldsymbol{1}\delta(t-t')$.
 This assumption is valid in the limit of vanishing persistence time of the active particle. Here, the multiplicative noise $\xchi(\rr)$ is implemented using the It\^{o}'s prescription which results in the following Langevin equation~\cite{gardiner1985handbook}
 \begin{align}\label{ito_langevin}
 &\!\!\!\!\!\!\dot{\rr} = \frac{1}{2} \nabla D_a(\rr) + \sqrt{2(D_t + D_a(\rr))}\boldsymbol{\mathcal{N}}\,,
 \end{align}
 where the term $\nabla D_a(r)/2$ is the noise-induced drift term and $\boldsymbol{\mathcal{N}}$ is Gaussian distributed with zero mean and time correlation $\langle \boldsymbol{\mathcal{N}}(t) \boldsymbol{\mathcal{N}}(t')\rangle= \boldsymbol{1}\delta(t-t')$.
We use Eq.~\eqref{ito_langevin} to derive Fokker-Planck equation for
 $P(\rr,t)$, defined as the probability density of finding an active particle at position $r$ at time $t$:

  \begin{align}\label{FPE}
    \frac{\partial}{\partial t}P(\rr,t) =\,\, &\nabla \cdot \left[\frac{1}{2} \left(\nabla D_a(\rr)\right)  P(\rr,t)\right] + \notag \\  
         &\nabla \cdot \left[( D_t + D_a(\rr)) \nabla P(\rr,t) \right].
 \end{align}


%
 
We note that $D_\text{a}(r)$ can be much larger than $D_t$ and hence, the diffusion of a particle may be governed predominantly by the activity. For noninteracting  particles, the enhanced diffusivity of active particles is reminiscent of Brownian particles at a vastly increased effective temperature~\cite{cates2013active}. In fact, Eq.~\eqref{FPE} describes a nonequilibrium process which breaks detailed balance and can be interpreted as describing a \sn{passive} system with spatially varying temperature. The Fokker-Planck equation obtained above is based on the Markovian process in Eq.~\eqref{ito_langevin}. However, even for the non Markovian process in  Eq.~\eqref{integrated_langevin}, there exist different schemes~\cite{fox1986functional,fox1986uniform,maggi2015multidimensional}, following which an approximate Fokker-Planck equation can be derived. These schemes yield a Fokker-Planck equation with first order correction in the persistence time of the particle~\cite{farage2015effective,sharma2017escape}. However, the correction is coupled to a potential term~\cite{farage2015effective,sharma2017escape}, which is not present in our model, and therefore the error in the white-noise approximation of $\chi(\rr)$ is of the order $\tau_a^2$.

We consider activity to be distributed spherically symmetric, continuously varying with distance $r$ from the centre as

\begin{align}\label{spatialv0}
v_0(r) = \frac{c}{r^{\alpha}}\left[\int_{r_0}^{R}dr 4\pi r^2\frac{1}{r^{\alpha}}\right]^{-1},
\end{align}
where the exponent $\alpha$ is varied to obtain different distributions. We consider $\alpha \geq -1$. The volume intergal of the activity is $c$ which can be interpreted as the `total activity' available in the spherical geometry between $r_0$ and $R$. The region $0<r\leq r_0$ may represent a source of fuel for the active particles~\cite{peng2015self} (see Fig.~\ref{schematic}). We assume that the activity, i.e., the self-propulsion speed of an active particle is proportional to the local concentration of the fuel. The total activity is, in this sense, proportional to the total amount of fuel present in the system. If one considers that the fuel source is emitting fuel at a constant rate which then diffuses isotropically in the surroundings, one obtains the steady state fuel distribution as $1/r$ which corresponds to $\alpha =1$. If the source emits a constant number of fuel particles per unit time which travel ballistically radially outwards, the corresponding steady state fuel is distributed as $1/r^2$. Since our focus in this study the effect of different distributions, we do not concern ourselves with the specific details of how a particular fuel distribution is obtained. In order to compare the effect of different distributions, we impose the constraint of fixed total activity (fuel) $c$ for all values of $\alpha$. 



\begin{figure}[t]
\includegraphics[width=0.9\columnwidth]{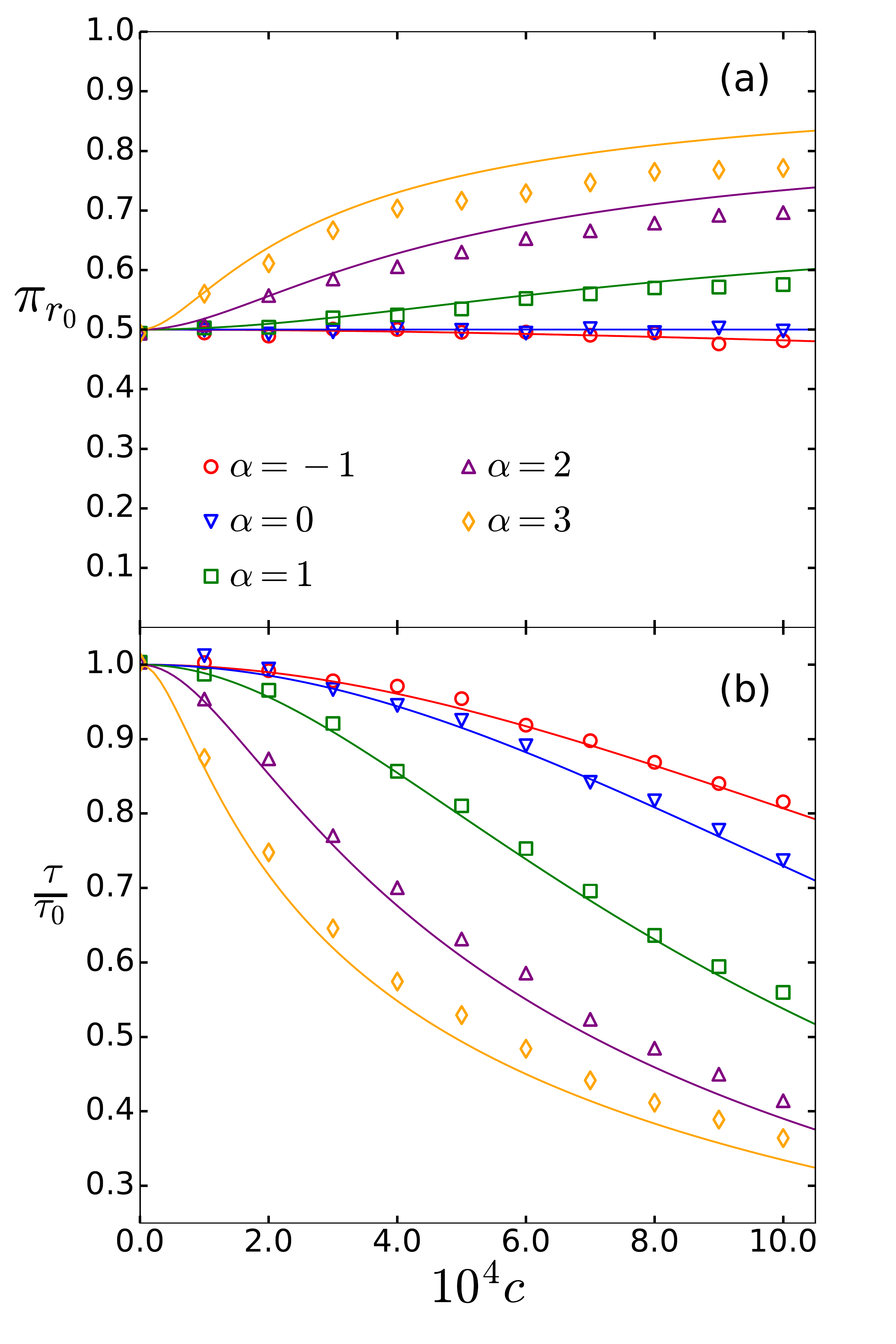}
		\caption{Target finding probability (a) and the mean first passage time to target (b) of active particles as a function of the total activity $c$. The activity profiles considered are of the form $v_0(r) \propto r^{-\alpha}$. Each profile is normalized such that the total activity is same for all values of $\alpha$. \sn{Simulation data are shown as symbols obtained from Brownian dynamics simulation of Eqs.~\eqref{full_langevin}.} 
The lines correspond to the theoretical prediction of Eq.~\eqref{exitprobability} in (a) and Eq.~\eqref{MFPT} in (b). The target finding probability $\pi_{r_0}(r)$ is calculated for $r=50/3$, which yields equiprobable exit from $r=r_0$ or $r=R$. $\pi_{r_0}(r)$ does not change when the system is uniformly active (circles). However, when the same amount of total activity is distributed such that it increases towards the target $r=r_0$, the probability is strongly biased. The mean first passage time to target ($r=r_0$) in (b) is calculated for a particle starting at $r=R$. It is normalized to its corresponding value $\tau_0$ in a passive system. Inhomogeneously distributed activity leads to a larger decrease in MFPT in comparison to a uniformly active system. 
}
		\label{pi_and_MFPT}
	\end{figure}

In Brownian dynamics simulations, we consider noninteracting particles with translational diffusion constant $D_t = 1/30$. The rotational diffusion constant $D_r = 1/2$. This corresponds to a quick rotation of the particle as compared to translation in the spirit of our approximation of short reorientation times made above. Based on the choice of parameters, the radius $a$ of the particle can be calculated using the Stokes-Einstein $D_{r}/D_t \sim a^{-2} \approx 1/4$. The total activity $c$ is a free parameter. The trajectory of each particle is generated by integrating the Langevin equations in Eq.~\eqref{full_langevin} using a time step $dt = 3\times10^{-3}\tau_D$, where $\tau_D = 1/D_t$ is the time scale of translational diffusion over a unit length. We fix the inner boundary of the spherical geometry as $r_0 = 10$. The outer boundary is fixed to $R = 50$. With these parameters the distance between the inner and outer boundary is much larger than the particle's diameter. 

\section{Target finding probability}\label{pseudochemotaxis}

We first consider target finding probability, $\pi_{r_0}(r)$, defined as the probability that a particle that is introduced at $r$ at time $t=0$ reaches the specified target, i.e., exits  through the inner boundary $r_0$ before it vanishes through the outer boundary. This is shown in Fig.~\ref{pi_and_MFPT}(a) as a function of the total activity $c$ and can be calculated from the Fokker-Planck Eq.~\eqref{FPE} as (Chapter 5 section 2 of \cite{gardiner1985handbook})

\begin{align}\label{exitprobability}
\pi_{r_0}(r) 
& = \frac{\int_{r}^Rdz\  z^{-2} \left(D_t + D_\text{a}(z)\right)^{-\frac{1}{2} }}{\int_{r_0}^Rdz\  z^{-2} \left(D_t +  D_\text{a}(z)\right)^{-\frac{1}{2} }}.
\end{align}
\
The scenarios considered are (i) the activity increases as one moves away from the target $v_0(r) \propto r$, corresponding to $\alpha = -1$, (ii) uniformly distributed activity as corresponding to $\alpha = 0$ and (iii) activity distributed  such that it increases towards the inner boundary $r_0$ (target) corresponding to $\alpha = 1,2,3$ (Eq.~\eqref{spatialv0}). 
The target finding probability does not change in the case of uniform activity, see Fig.~\ref{pi_and_MFPT}(a). 
It is only when the activity is inhomogeneously distributed, the probability is strongly biased to reach the target $r_0$ located at higher activities.
As can be seen in Fig.~\ref{pi_and_MFPT}(a), the theoretical predictions are in good agreement with the simulation data. The starting location $r=50/3$ is chosen as it corresponds to an equally likely exit from either of the two boundaries in a passive or a uniformly active system. We note that the qualitative behavior remains the same for any other starting location, i.e., probabilities to exit from either of the boundaries do not change in presence of uniform activity whereas in the case of inhomogeneous activity, the probability increases at the end where the activity increases. Although, here we consider smoothly distributed activity as in Eq.~\eqref{spatialv0}, the same qualitative behavior is obtained for piecewise distributed activity. For instance, if the activity is assumed to be uniform between $r_0$ and $R$ except a step-like larger activity of arbitrary length anywhere between $r_0$ and $r$, the probability of escaping through the target boundary increases.

\section{Mean first passage time to target}\label{Meanpassage}

The MFPT of an active particle starting at the outer boundary $r=R$  to reach the target at $r=r_0$ is shown in Fig.~\ref{pi_and_MFPT}(b). Considering $r=R$ as a reflecting boundary, MFPT to target of a particle, $\tau(r)$, is the average time taken by a particle starting at $r$ to reach the target $r=r_0$. This can be calculated from the Fokker-Planck Eq.~\eqref{FPE} as (Chapter 5 section 2 of \cite{gardiner1985handbook})

\begin{align}\label{MFPT}
	\tau(r) 
	& = \int_{r_0}^rdz\  \frac{z^{-2}}{\sqrt{D_t + D_\text{a}(z)}}
	\int_z^Rdy\
	\frac{y^{2}}{\sqrt{D_t +D_\text{a}(y)}}.
\end{align}
We normalize the MFPT with its corresponding value in a passive system ($\tau_0$). Increasing activity always decreases the MFPT. However, the decrease in MFPT is much more pronounced when the activity is spatially distributed such that it increases towards the target. As can be seen in Fig.~\ref{pi_and_MFPT}(b), the theoretical predictions are in good agreement with the simulation data. In our coarse-grained approach, it can be easily seen from Eq.~\eqref{FPE} that a uniformly active system is equivalent to a passive system with an effective diffusion constant $(D_t + D_\text{a})$. The decrease in the MFPT can thus be simply attributed to the increased diffusivity of the particle. However, when the activity is inhomogeneously distributed, the decrease in MFPT is more pronounced.



\begin{figure}[t]
		\includegraphics[width=\columnwidth]{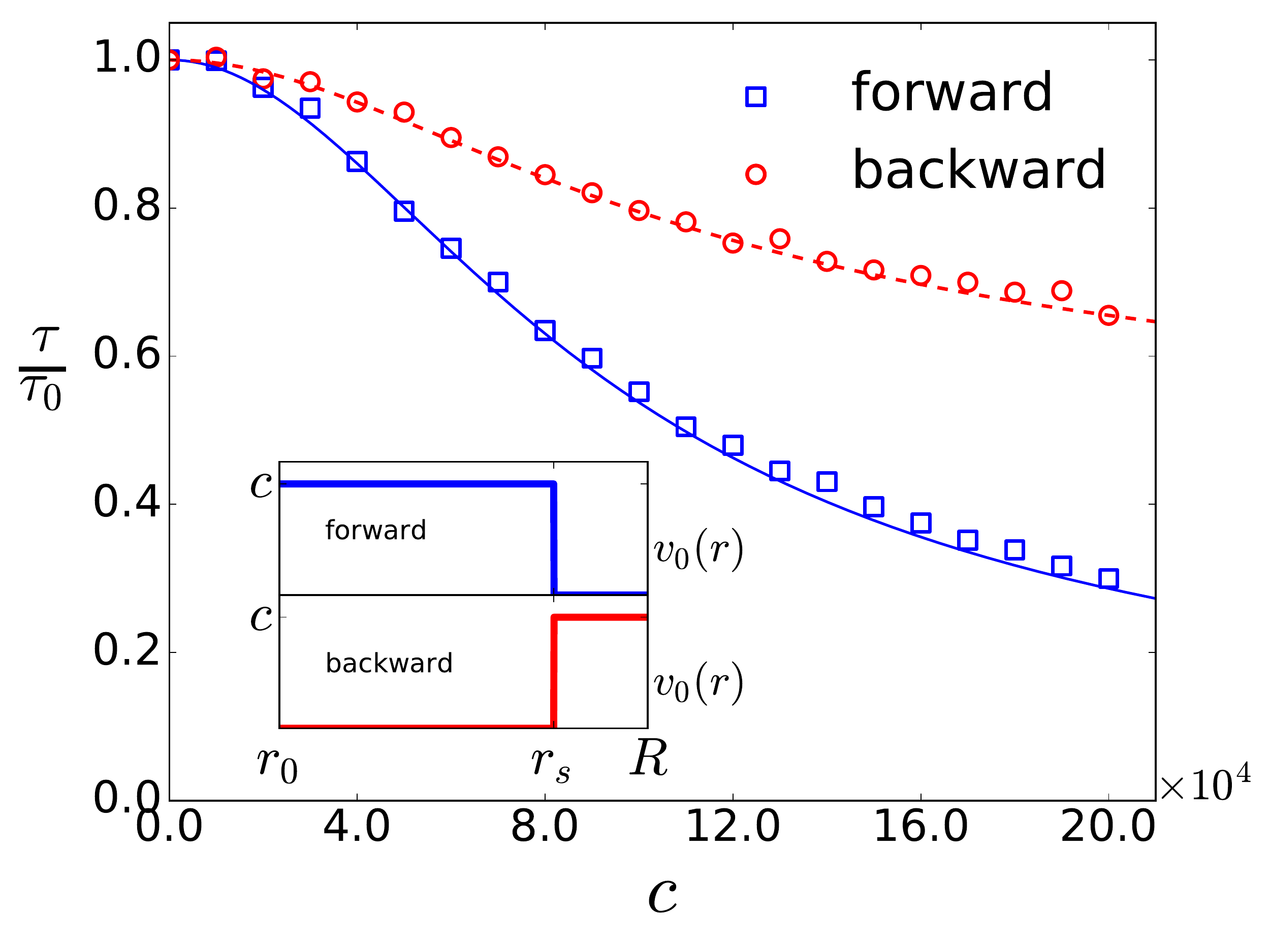}
		\caption{Mean first passage time to target for stepwise uniform activity as shown in the inset. Symbols denote data from Brownian dynamics simulations and the lines to the theoretical predictions of Eq.~\eqref{MFPT}. The forward scenario corresponds to the active region close to the target whereas in the backward scenario, the active region is away from the target towards the outer boundary. Both scenarios have the same amount of total activity. The forward scenario yields a much faster passage to the target.}
		\label{step}
\end{figure}

As in the case of target finding probability, the spatial distribution of the activity strongly influences the MFPT. A particularly simple but instructive case that illustrates the role of the spatial distribution of activity can be constructed as follows. We consider two scenarios, called the forward and the backward scenario. In the forward scenario, the region $r_0<r<r_s$ is uniformly active and $r_s<r<R$ passive. In the backward scenario, the situation is reversed with the active region becoming passive and vice versa. The intermediate distance $r_s = \sqrt[3]{(r_0^3 + R^3)/2}$ is chosen such that the total activity in both scenarios is the same (see inset of Fig.~\ref{step}). The activity gradient is everywhere zero except at $r=r_s$ implying that the noise-induced drift of the particle occurs only at $r=r_s$ towards $r_0$ in the forward scenario and $R$ in the backward scenario. 
In Fig.~\ref{step}, we plot the MFPT as a function of the total activity $c$ for the forward ($\tau_f(R)$) and backward ($\tau_b(R)$) scenarios. Clearly, the forward scenario yields a much faster passage to the target. One can show in a straightforward calculation that the difference between the MFPTs of the backward and the forward scenario, $\delta \tau = \tau_b(R) - \tau_f(R)$ is given as 
\begin{align}
\delta \tau =\frac{D_\text{a}}{6(D_t + D_\text{a})}\left[\frac{2r_s^3}{r_0}- \frac{2R^3}{r_s} + 3R^2 - 4r_s^2 + r_0^2  \right],
\end{align}
where $D_\text{a} = c^2/(6D_r)$. It can be easily shown that $\delta \tau$ is always positive. This simple case serves to illustrate the strong influence of the spatial distribution of activity on the MFPT. With the active region closer to the absorbing boundary, the MFPT in the forward scenario is significantly smaller than in the backward scenario.

The agreement between the theoretical predictions and the simulations degrades with increasing $\alpha$ as can be seen in Fig.~\ref{pi_and_MFPT}. The theoretical description based on Eqs.~\eqref{integrated_langevin} and~\eqref{ito_langevin} ignores the coupling between fluctuations in orientation and positional degrees of freedom. With increasing $\alpha$, the activity increases near the target and the position of the particle can change significantly during orientational relaxation. Ignoring this coupling between orientation and position is the main reason for the  disagreement between theory and simulations.


\begin{figure}[t]
		\includegraphics[width=\columnwidth]{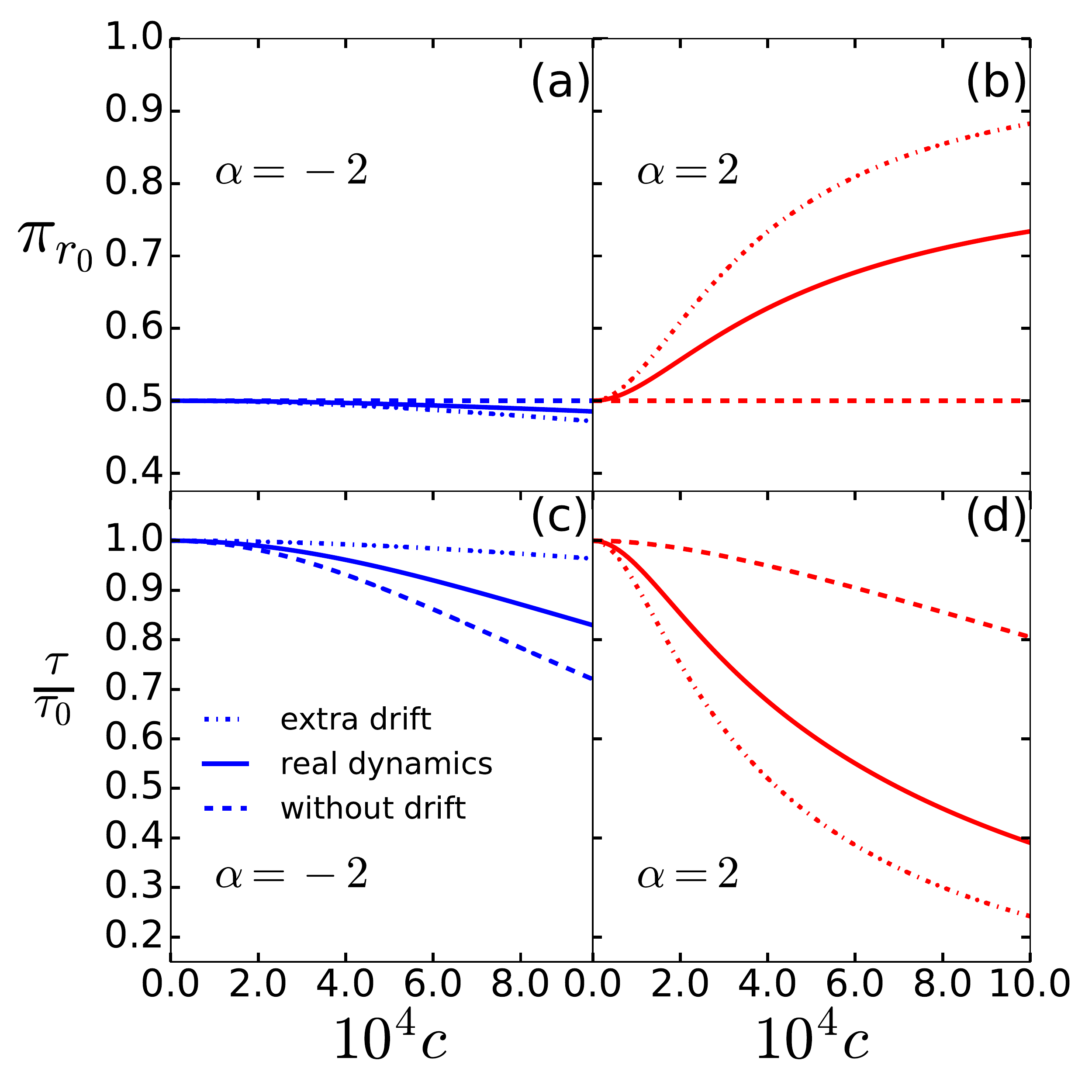}
		\caption{Target finding probability (a,b) and MFPT (c,d) as function of total activity $c$. The solid lines are calculated using Eq.~\eqref{FPE} corresponding to the `real' dynamics of the particle according to Eq.~\eqref{ito_langevin}. The dashed lines are calculated using Eq.~\eqref{FPE2} and correspond to the fictitious dynamics without drift. The dash-dotted lines are calculated using Eq.~\eqref{FPE3} and correspond to the fictitious dynamics with extra drift. $\alpha = -2$ corresponds to the $r^2$ distributed activity with activity gradients pointing away from the target and $\alpha = 2$ to the $r^{-2}$ distributed activity (same as Eq.~\eqref{spatialv0}) with activity gradients pointing towards the target. In the case of no drift, the target finding probability remains constant. On switching off the drift, the MFPT becomes larger for the $r^{-2}$ distributed activity and smaller for $r^2$ distributed activity. In the case of extra drift, the target finding probability becomes larger for the $r^{-2}$ distributed activity and smaller for the for $r^2$ distributed activity. The MFPT becomes smaller for the $r^{-2}$ distributed activity and larger for $r^2$ distributed activity. Clearly, the noise-induced drift facilitates the passage to the target for $r^{-2}$ distributed activity whereas the opposite holds true for $r^2$  distribution. 
		}
		\label{fig_drift}
	\end{figure}

\section{Role of activity gradients}
There are two contributing factors to the increase in the target finding probability: (a) more activity located between the starting location and the target than outside and (b) noise-induced drift due to the spatial variation of activity. We consider these two factors individually by first considering a scenario in which more activity is located between the starting location $r$ and the target $r_0$ than outside. For instance, if the activity is assumed to be uniform between $r_0$ and $R$ except a step-like larger activity between $r_0$ and $r$. This scenario can be mapped to a system with larger temperature between $r$ and $r_0$ than outside this region. A hot region is covered faster by diffusion than a cold region leading to an increased target finding probability. Next we consider the contribution of the noise-induced drift for continuously distributed activity as in Eq.~\eqref{spatialv0}. As we show below, the noise-induced drift is also the main reason for the decrease in MFPT in an inhomogeneous active system. The noise-induced drift arises due to spatial variation of activity (Eq.~\eqref{ito_langevin}). In order to find out the effect of the noise-induced drift, we consider fictitious dynamics of the particle by switching off the drift term in Eq.~\eqref{ito_langevin}. In this case, the Langevin dynamics correspond to $\dot{\rr} =  \sqrt{2(D_t + D_a(\rr))}\boldsymbol{\mathcal{N}}$, from which the following Fokker-Planck equation is obtained:
  \begin{align}\label{FPE2}
    \frac{\partial}{\partial t}P(\rr,t) = \nabla \cdot \left[\nabla (\left( D_t + D_a(\rr) P(\rr,t)\right)  \right].
 \end{align}
Using Eq.~\eqref{FPE2} we analytically calculate the target finding probability and MFPT which is shown in Fig.~\ref{fig_drift}. We consider two inhomogeneous distributions (i) decaying as $r^{-2}$ away from the target (same as Eq.~\eqref{spatialv0}) and (ii) increasing as $r^2$ away from the target towards the outer boundary. The latter corresponds to activity gradient pointing away from the target whereas the latter corresponds to activity gradients pointing towards the target. We find that on neglecting the noise-induced drift term, the target finding probability remains $0.5$ independent of the total activity $c$ for both the distributions. Interestingly, considering MFPT, we find that on neglecting the noise-induced drift term, it becomes significantly larger for the $r^{-2}$ distribution. When the activity is distributed as $r^2$, i.e., with activity gradients pointing away from the target, removing the noise-induced drift leads to a significant decrease in the MFPT. It thus follows that whereas the noise-induced drift facilitates the passage to the target for $r^{-2}$ distributed activity, the opposite holds true for $r^2$  distribution. 

Rather than removing the drift term, if one adds $\nabla D_a(\rr)/2 $ to the Langevin equation~\eqref{ito_langevin}, such that the dynamics correspond to $\dot{\rr} = \nabla D_a(\rr) + \sqrt{2(D_t + D_a(\rr))}\boldsymbol{\mathcal{N}}$, one obtains the corresponding Fokker-Planck equation as
  \begin{align}\label{FPE3}
    \frac{\partial}{\partial t}P(\rr,t) = \nabla \cdot \left[( D_t + D_a(\rr) \nabla P(\rr,t)  \right].
 \end{align}
This equation corresponds to a fictitious system with space-dependent diffusion coefficient. We have also performed analytical calculations corresponding to this case for the two activity distributions as mentioned above (see Fig.~\ref{fig_drift}). We find that the addition of noise-induced drift leads to an even stronger decrease in MFPT and a larger increase in the target finding probability than that from the Langevin equation~\eqref{ito_langevin}. It follows that the noise-induced drift strongly affects both target finding probability and the MFPT in an inhomogeneous active system.


\section{Pseudochemotaxis}
The increase in likelihood of escaping through the target boundary is reminiscent of the chemotaxis phenomenon~\cite{berg2008coli}. Chemotaxis is a fundamental sensory mechanism by which bacteria and other single- or multicellular organisms monitor the concentration gradients of specific chemicals, translating the information into motion either uphill or downhill to the gradient. The increased likelihood of escaping from the inner boundary, where activity increases, can be likened to an active particle climbing up the fuel gradient. This chemotactic behavior has been recently realised in experiments on supramolecular nanomotors which climb up the hydrogen peroxide concentration gradient~\cite{peng2015self}. 

Considering that the stationary distribution of active particles in an inhomogeneous activity profile tends to accumulate in the low activity region, the chemotactic behavior of active particles appears paradoxical. Recently, Ghosh \emph{et. al}~\cite{ghosh2015pseudochemotactic} addressed this paradox by emphasizing the distinction between the dynamical and stationary behavior of inhomogeneous active systems. The stationary distribution is obtained under the assumption that the active particle is trapped between two reflecting boundaries. The target finding probability, in contrast, is the likelihood of reaching a target boundary. In a stationary scenario, the drift of the particle towards the end where activity increases, is a dynamical effect and does not impact the stationary distribution.

\begin{figure}[t]
		\includegraphics[width=\columnwidth]{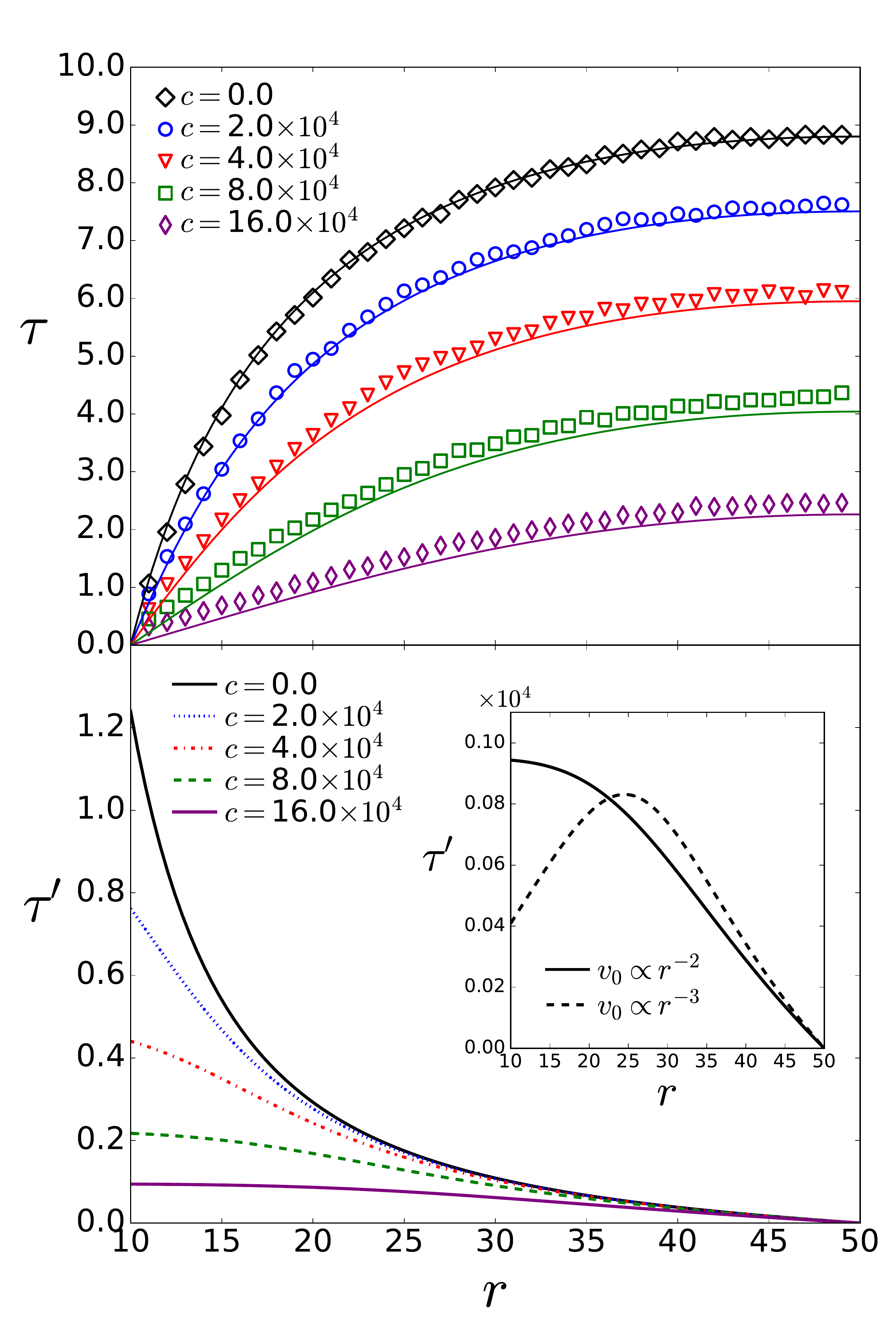}
		\caption{(a) Mean first passage time to target as a function of the starting location $r$ for different values of $c$. The activity is distributed according to Eq.~\eqref{spatialv0}. Symbols denote data from Brownian dynamics simulations and the lines to the theoretical predictions of Eq.~\eqref{MFPT}. The rate of change of MFPT with $r$ is calculated using Eq.~\eqref{MFPT} and is shown in (b). The inset of (b) replots $\tau'(r)$ corresponding to $c=16\times10^4$ to better visualize the nonmonotonic behavior. When the same amount of total activity is distributed as $1/r^3$, $\tau'(r)$ exhibits a maximum which moves aways from $r_0$ towards $R$ with increasing $c$.
		}
		\label{geometry}
	\end{figure}

\section{Phase-space effect}			
	
We consider how the MFPT changes as a function of the distance $r$ from the target $r=r_0$. In Fig.~\ref{geometry}(a) we plot the MFPT as a function of $r$ for different values of $c$ when activity is distributed inhomogeneously as in Eq.~\eqref{spatialv0}. As can be seen in the Fig.~\ref{geometry}(a), the MFPT decreases monotonically with $c$ for any given $r$. For a fixed $c$, the MFPT, as expected, increases with increasing $r$. This seems to be a trivial observation as one expects $\tau(r)$ to increase with increasing distance from the target.
 However, there is a subtle geometrical aspect that becomes evident when one considers $\tau'(r)$, the rate of change of $\tau(r)$ with respect to $r$ as shown in Fig.~\ref{geometry}(b). For a passive system ($c=0$), $\tau'(r)$ is largest for $r=r_0$ and then decreases monotonically with increasing $r$. However, in an active system, $\tau'(r)$ shows a qualitatively different behavior with $r$. With increasing $c$, the decay of the $\tau'(r)$ near $r_0$ becomes increasingly slow, becoming nearly flat.

The qualitative change in the behavior of $\tau'(r)$ is due to an underlying competing effect between the available phase-space and activity which can be understood as following. For a particle introduced at $r$ such that $r_0<r<R$, the passage to $r=r_0$ includes trajectories spanning the volume between $r$ and the reflecting boundary at $R$. The smaller $r-r_0$ is, the larger is this extra volume (phase-space) between $r$ and $R$ in which the particle can wander before being absorbed at $r=r_0$. This phase-space effect gives rise to the observed strong decrease of $\tau'(r)$ in a passive system. However, in an inhomogeneous active system, there is an additional competing effect due to the activity which tends to prevent the excursion of the particle away from $r_0$. With increasing $c$, the noise-induced drift becomes increasingly important and counters the wandering of the particle. These two competing effects give rise to the observed slower decay of $\tau'(r)$. We emphasize that even for $\tau'(r)$, the emergence of this subtle competing effect between the available phase-space and activity is due to the inhomogeneous distribution of the activity. In a system with uniform activity, $\tau'(r)$ decays strongly away from the absorbing boundary as for the passive system. The behavior of $\tau'(r)$ is strongly dependent on the activity profile. 

Interestingly, when the activity decays faster than $1/r^2$ in three-dimensions,  $\tau'(r)$ becomes nonmonotonic. In the inset of Fig.~\ref{geometry}(b), we plot $\tau'(r)$ for the activity distributed as $1/r^3$ corresponding to a total activity of $c=16\times10^4$. With this choice of activity profile, one obtains a maximum in $\tau'(r)$ which shifts away from $r_0$ towards $R$ with increasing $c$. Such rapidly decaying activity profile would naturally arise in systems in which the fuel molecules injected by the source, have a finite lifetime. Fuel molecules may bind to species other than the active particles resulting in an attenuation of freely available fuel. An activity profile that decays faster than $r^{-(d-1)}$ for $d\geq2 $ does not only compensate for the competing phase-space effect but dominates it near $r_0$ for larger $c$ giving rise to the observed nonmonotonicity of $\tau'(r)$. This can be easily demonstrated quantitatively by differentiating Eq.~\eqref{MFPT} twice with respect to $r$ for the chosen activity profile. We note that the phase-space related effects as discussed here are absent inside one-dimensional systems such as linear channels, but are of significance in majority of natural scenarios.

Active particles exhibit non-isotropic distribution of their orientation vectors in presence of an activity gradient~\cite{sharma2017brownian}. In the current setup, however, these gradients are insufficiently steep to generate significant effects on the walker's orientations. In fact, our Fokker-Planck equation
reproduces the computational findings in complete absence of any orientational inhomogeneities, which is also in agreement with earlier study by Ghosh \emph{et al.}~\cite{ghosh2015pseudochemotactic}. It is therefore safe to conclude that orientational effects are of no relevance
for the target finding probability and MFPT in the context of our work. In our simulations of active particles we find that $\tau(r)$ exhibits a small but finite discontinuity at $r=r_0$. This is a consequence of the finite persistence time of active particles which is ignored in our theoretical approach. Nevertheless, except in the immediate neighborhood of $r_0$, the theoretical predictions are in very good agreement with the simulations. 



\section{Conclusions and outlook}

We studied the dynamical properties of noninteracting active particles in an inhomogeneous activity profile. Using Brownian dynamics simulations, we calculated the probability to reach a fixed target and the mean first passage time to the target of an active particle. We showed that both these quantities are strongly dependent on the spatial distribution of the activity. When the activity is distributed such that high-activity zone is located between the target and the starting location, the target finding probability is increased and the passage time is decreased in comparison to a uniformly active system. Moreover, for a continuously distributed profile, the activity gradient results in a drift of active particle up the gradient bearing resemblance to chemotaxis~\cite{ghosh2015pseudochemotactic,geiseler2016chemotaxis}. We further showed that inhomogeneous activity can give rise to subtle effects such as the nonmonotonic behavior of $\tau'(r)$, which are absent in uniformly active systems or linear channels.


We found that the insights gained from steady-state inhomogeneous active systems, such as preferential accumulation in the low-activity regions and orientation of particles antiparallel to the activity gradient, cannot be used to understand the dynamical properties of an inhomogeneous active system. \sn{The noise-induced drift emerges naturally in a system with spatially varying noise and it points in the direction of the activity gradient. As a consequence, a particle starting anywhere in the system drifts towards higher activity. The particle moves increasingly faster as it gets closer to the target. The drift aids the passage of the particle to the target giving rise to the observed increase in target finding probability. However, this does not mean that the particle has a larger residence time in the high activity regions. If the particle is reflected from the target, it can move into low activity regions where it resides for a longer time than in the high activity regions.}


Finally, considering activity profiles in general the following interesting questions arise:  Which activity profile, for a given total activity $c$, yields the minimum mean first passage time? Which profile yields the maximum target finding probability? In a very recent study~\cite{liebchen2018modelling} on chemotaxis, the authors have considered a source emitting a chemical signal which develops a spatio-temporal distribution. It will be very interesting to extend our study to such systems  in which the activity profile is dependent both on space and time. In particular, what will be the target finding probability in this scenario, for both static and moving target? In the near future, we will include interaction potential between the source ($r_0$) and active particle. It will be particularly interesting to investigate under what conditions an active particle exhibits chemotactical behavior when it interacts via a repulsive interaction with the source of fuel.

%
%

\end{document}